\documentclass[preprint,numbers]{aastex}

\usepackage{amsmath}
\usepackage{amsfonts}
\usepackage{amssymb}
\usepackage{graphicx}
\usepackage{epsfig}

\begin{document}

\title{CF2 White Paper: Status and Prospects of The VERITAS Indirect Dark Matter Detection Program}

\author{
A.~W.~Smith\altaffilmark{1*},
R.~Bird\altaffilmark{2}
J.~Buckley\altaffilmark{3},
K.~Byrum\altaffilmark{4}
J.~Finley\altaffilmark{5},
N.~Galante\altaffilmark{6}
A.~Geringer-Sameth\altaffilmark{7},
D.~Hanna\altaffilmark{8}
J.~Holder\altaffilmark{9},
D.~Kieda\altaffilmark{1},
S.~Koushiappas\altaffilmark{7},
R.A.~Ong\altaffilmark{10}
D.~Staszak\altaffilmark{8}
B.~Zitzer\altaffilmark{4}
}

\altaffiltext{*}{aw.smith@utah.edu}
\altaffiltext{1}{Department of Physics and Astronomy, University of Utah, Salt Lake City, UT 84112, USA}
\altaffiltext{2}{School of Physics, University College Dublin, Belfield, Dublin 4, Ireland}
\altaffiltext{3}{Department of Physics, Washington University, St. Louis, MO 63130, USA}
\altaffiltext{4}{Argonne National Laboratory, 9700 S. Cass Avenue, Argonne, IL 60439, USA}
\altaffiltext{5}{Department of Physics, Purdue University, West Lafayette, IN 47907, USA }
\altaffiltext{6}{Fred Lawrence Whipple Observatory, Harvard-Smithsonian Center for Astrophysics, Amado, AZ 85645, USA.  }
\altaffiltext{7}{Department of Physics, Brown University, 182 Hope Street, Providence, RI 02912, USA }
\altaffiltext{8}{Physics Department, McGill University, 3600 University Street, Montreal, QC, H3A 2T8}
\altaffiltext{9}{Department of Physics and Astronomy and the Bartol Research Institute, University of Delaware, Newark, DE 19716, USA}
\altaffiltext{10}{Department of Physics and Astronomy, University of California, Los Angeles, CA 90095, USA}


 
\section{Introduction}

The field of ground based gamma-ray astronomy has seen an dramatic increase in the scale and breadth of its contribution to astrophysics over the last decade. The previous generation of imaging atmospheric Cherenkov telescopes (IACTs) such as the Whipple 10m \citep{Whipple} and HEGRA \citep{HEGRA} incontrovertibly demonstrated the feasibility and scientific value of detecting cosmic TeV sources from the ground. However, these detections were limited to a handful of sources (mostly active galactic nuclei), due to the limitations of the technology at hand. The current generation of IACTs such as VERITAS, HESS, and MAGIC \citep{VERITAS,HESS, MAGIC} have impressively progressed the field by detecting over 140 cosmic TeV sources (many extended) across a wide range of source classes including supernova remnants, X-ray binaries, radio galaxies, and AGN at redshifts previously thought out of reach to the technique.

One of the more exciting prospects for the current generation of IACTs is to use their capabilities to search for signatures of particle dark matter. In the standard $\Lambda$-CDM paradigm, dark matter appears as a cold, non-baryonic, weakly interacting massive particle (WIMP) with a relic density of  $\Omega_{CDM}$h$^{2}$ = 0.1109 $\pm$0.0056 \citep{Komatsu}. Enthusiasm for this model has been spurred on by its success in reproducing large scale structures in simulations; additionally, many reasonable candidates for the WIMP naturally occur in beyond Standard Model (SM) theories such as Supersymmetry (SUSY) and models incorporating large extra dimensions \citep{SUSY,KK}. The WIMP in these models typically either decays or self-annihilates into normal SM particles, thus allowing for indirect detection of dark matter through their decay or annihilation products. In the context of TeV astronomy, this approach to indirect dark matter detection has typically invoked the lightest SUSY particle, or neutralino ($\chi$), as a WIMP candidate as its self annihilation produces gamma-ray signatures in the GeV-TeV energy range. This annihilation can produce gamma rays either through intermediary annihilation channels (such as quark or lepton states), or directly into gamma rays with photon energies identical to the rest energy of $\chi$ ( $\chi\chi\rightarrow\gamma\gamma$ ) , or energies near to the  $\chi$ rest energy ($\chi\chi\rightarrow\gamma Z/h$). Observationally, the former would be represented by a spectrum of gamma rays with a sharp cutoff at the neutralino mass (continuum emission), whereas the latter would present as gamma-ray lines within the continuum emission. While the line signals would be unambiguously herald the presence of self-annihilating DM, the cross section associated with a gamma-ray line is expected to be loop-suppressed and a factor of 10$^{4}$ weaker than the continuum emission. 
\begin{figure}[t]
\begin{center}
   \includegraphics[width=\textwidth,height=50mm]{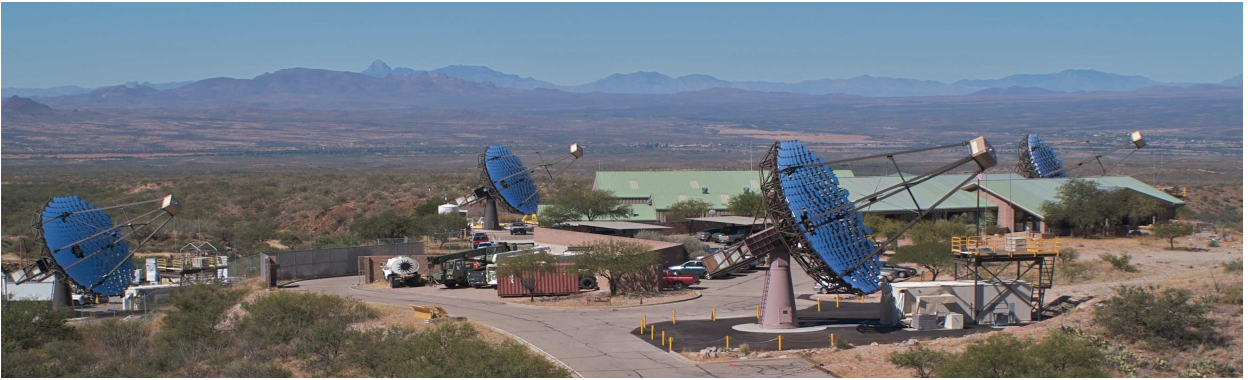}
   \caption{The VERITAS Imaging Atmospheric Cherenkov Array, located approximately 30 miles south of Tucson, AZ at the base of Mt. Hopkins (1.3km a.s.l.). }

\end{center}
\end{figure}

  Since DM interacts gravitationally, large concentrations of mass would ostensibly provide large over-densities of DM as well. Therefore, the observation of a TeV gamma-ray signal from one of these concentrations would provide evidence for self-annihilating DM. Gamma rays in this energy range are also produced by astrophysical objects such as pulsars, supernova remnants, and the interacting winds of massive stars; the typical approach has been to observe over-dense mass regions with both high mass-to-light ratios, as well as a presumed lack of dominant, conventional sources of TeV radiation. DM targets such as dwarf spheroidal galaxies (dSphs) and galaxy clusters fulfill both of these requirements and have served as natural targets for observations with IACTs in the search for DM annihilation. Added to this list is the Galactic center of the Milky Way (GC): although it is a "busy" TeV region of the sky, its proximity to us as well as its presumed DM content merit the use of sophisticated analysis techniques in order to attempt to separate the contribution of conventional TeV sources from that which could be explained by DM annihilation. 

This approach to searching for DM has been employed by the VERITAS IACT array \citep{VERITAS}. Located in Southern Arizona, VERITAS is comprised of four, 12 meter diameter Davies-Cotton optical reflectors which focus light from gamma-ray air showers onto four 499 pixel photomultiplier tube( PMT) cameras. The array, with a total field of view of 3.5$^{\circ}$, is sensitive in the range of  100 GeV  to 50 TeV, easily covering a large range of parameter space favored by the SUSY models discussed previously. Since the commissioning of the array in 2007, VERITAS has accrued approximately 350 hours of observations on a range of putative dark matter targets including dSphs, galaxy clusters and the GC. The analysis of these observations is ongoing, however, the results already published from subsets of the observations have resulted in some of the strongest DM constraints available from indirect searches. In this white paper we summarize these results, focusing on the constraints obtained from dSph and GC observations and highlighting the improvements offered by new data analysis techniques (such as source stacking) which are quickly reaching maturity. As observations for the VERITAS DM program are ongoing and will continue through the operational lifetime of the array, these constraints will only be improved with time. The continuing accrual of data will be strongly augmented by the recent upgrade of VERITAS, which has replaced all the existing PMTs with high quantum efficiency units, offering enhanced sensitivity, especially below 200 GeV. We argue that the plan for a continued and extensive dark matter search program with VERITAS offers one of the strongest avenues available for the possible detection of WIMP dark matter and, in the absence of detection, can severely constrain many conservative SUSY models.

\section{Current Results and Forecasts}

Following \citet{Wood}, the differential flux of gamma rays resulting from neutralino annihilations from a given source can be expressed as:

\[\displaystyle {\frac{d\Phi_{\gamma}}{dE}(\Delta\Omega, E) = \frac{\langle \sigma v\rangle }{8\pi m_{\chi}^{2}}\frac{dN_{\gamma}}{dE}\times J(\Delta\Omega)}\]

where  $\langle \sigma v\rangle $ is the thermally averaged neutralino self-annhilation cross section, m$_{\chi}$ is the mass of the neutralino, and dN$_{\gamma}$/dE represents the gamma-ray differential energy spectrum for annihilation. For the work shown here,  we model  dN$_{\gamma}$/dE for several different annihilation channels including both pure quark and lepton intermediary states. The factor $J(\Delta\Omega)$ is a factor which represents the DM content within the solid angle subtended by the observations, and it is therefore strongly a function of the overall astrophysical DM density profile. The choice of DM profiles is one of the chief uncertainties involved in producing the relevant constraints; the determination of the actual astrophysical profile of DM is an extremely active area of research (e.g. \citet{Walker}). 

In the event of a null detection of a gamma-ray excess from a putative DM source (the case thus far), a flux upper limit on the number of counts coming from DM annihilation can be produced via any number of statistical methods. For the works presented here, we have chosen the Rolke method of producing flux upper limits \citep{Rolke}. Once a flux upper limit is obtained, this can be combined with a putative DM density profile and annihilation spectrum to produce (after folding in the solid angle of the detector in question) a constraint on $\langle \sigma v\rangle $.

\begin{figure}[t]
\begin{center}
   \includegraphics[width=\textwidth,height=60mm]{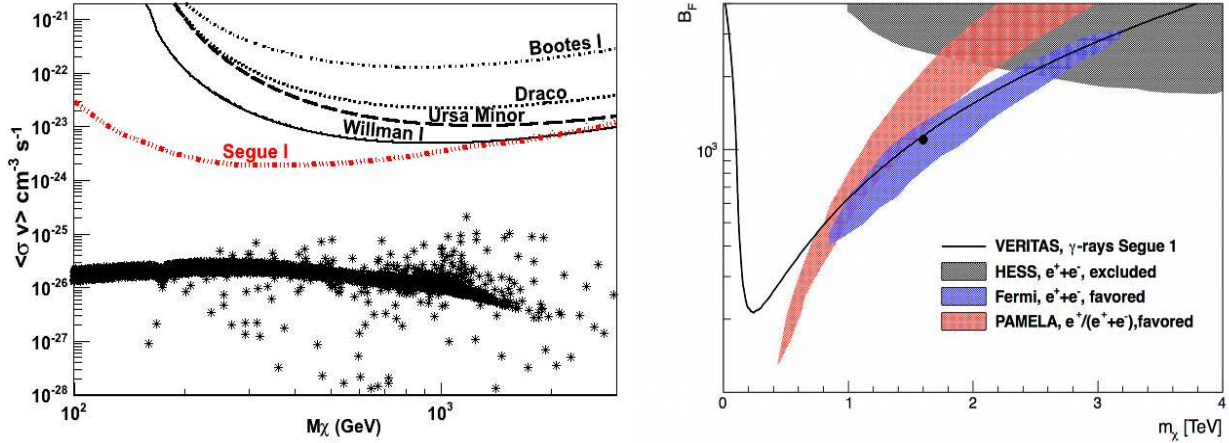}
   \caption{Results from the VERITAS dSph observations. In the left figure, the black curves represent a $\sim$10-15 hour observation on the dSphs Draco, Ursa Minor, Bootes I, and Willman I \citep{Bobs}. The red curve represents a deep observation ($\sim$ 50 hours) on Segue I \citep{Segue}. On the right are shown the VERITAS, HESS, and Fermi constraints on the boost factor required to explain the PAMELA positron fraction by a leptophillic dark matter model. }

\end{center}
\end{figure}

\subsection{Dwarf Spheroidal Galaxies}

While the observations of dSphs with VERITAS are ongoing, the publication of preliminary results from subsets of these observations have been of key importance and well received within the particle astrophysics community. The first VERITAS dSph results \citep{Bobs} were composed of observations of the dSph galaxies Draco, Ursa Minor, Bootes I, and Willman I (see Figure 2). This result was based on a relatively small observational exposure (10-15 hours per source) and was significantly improved by the result of \citet{Segue} in which over 50 hours of observation on the dSph Segue I was presented (Figure 2). While these limits do not constrain the most conservative realizations of minimal SUSY, the VERITAS dSphs limits (in particular the Segue I limits) provide much stronger constraints on alternative models of SUSY \citep{Segue} (for example, invoking a Sommerfeld enhancement or a scalar mediator).  Models with a kinetically enhanced cross section are already constrained by current VERITAS observations, and may be all but excluded over the next few years of VERITAS observations. 

In addition, VERITAS dSph observations have now provided strong constraints on models of dark matter annihilation invoked to reproduce the PAMELA positron excess \citep{PAMELA}. In these models, DM annihilates exclusively into $\mu^{+}\mu^{-}$ (leptophillic models). In order for these models to explain the PAMELA excess, a boost factor B$_{F}$ is required (both astrophysical and particle physics boosts are convolved into B$_{F}$). The VERITAS Segue I observations places strong constraints on B$_{F}$, limiting the allowed parameter space that can be utilized to explain the PAMELA excess within a DM framework (see Figure 2). These results are especially relevant in light of the recent results from the AMS experiment \citep{AMS}. As with the constraints on $\langle \sigma v\rangle $, these limits will improve with additional observations.

\subsection{The Galactic Center}
\begin{figure}[t]
\begin{center}
   \includegraphics[width=\textwidth,height=60mm]{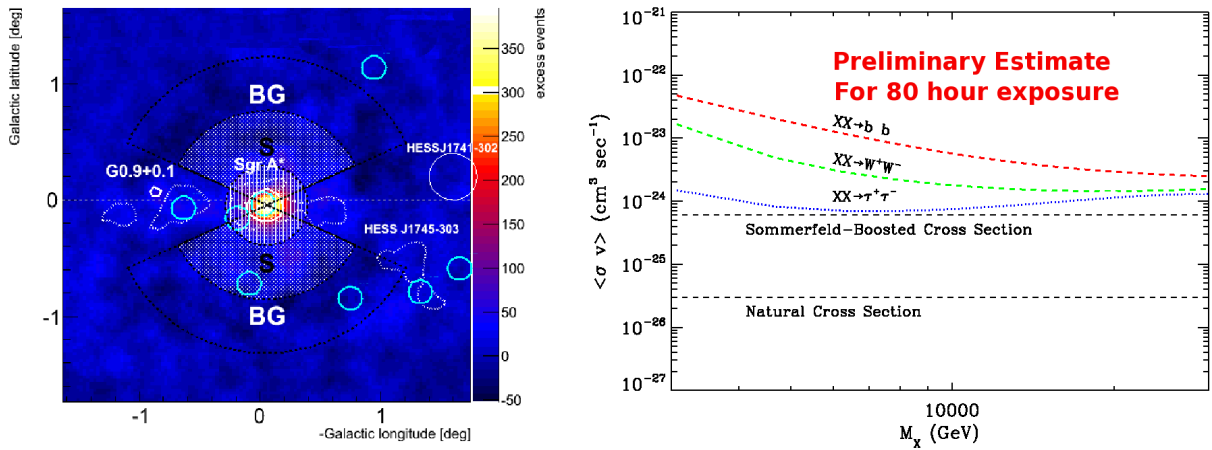}
   \caption{The VERITAS sky map of the GC region showing the signal and background regions to be used for DM analysis (left). Also shown (right) is the prediction for the VERITAS upper limits derived from roughly 80 hours of GC observations.}

\end{center}
\end{figure}
The center of our galaxy harbors a supermassive (4$\times$10$^{6}$ M$_{\odot}$) black hole (SMBH) coincident with the radio source Sgr A*. The Sgr A* region is a conventional source of both GeV and TeV gamma rays \citep{HESSGC,VERGC}, resulting from either relativistic protons accelerated in the vicinity of the SMBH or from the diffusion of the protons into the interstellar medium \citep{GCemission1, GCemission2}. In addition to the conventional sources of VHE gamma rays originating from the GC, this region is also expected to be the closest, dense region of DM accumulation. Therefore, despite its large background of known gamma-ray sources, the GC is a very important target for indirect DM detection searches. Care must be taken in the analysis to mask the normal TeV emission from the GC and leave the putative DM annihilation signal remaining (see Figure 3, left). VERITAS has currently accrued approximately 70 hours of livetime observations on the Sgr A* region, including ``OFF" source pointings to allow for better background determination. Due to the position in the sky of the GC for VERITAS, the observations of the GC have been carried out at large angles ($\>$ 50$^{\circ}$) from the zenith. This has resulted in an increased sensitivity to higher energy TeV radiation for these observations at the cost of an increased energy threshold. While an increased energy threshold for a given observation is normally not desirable in TeV astronomy, in this case, the increased zenith angles of the GC observations  allow VERITAS  to provide stronger constraints on the annihilation of DM coming from higher mass neutralinos; these constraints are not available from any other experiment (see Figure 3, right). This may be particularly important in light of the recent LHC results pointing to the existence of a 130 GeV range Higgs particle \citep{Higgs}. The relatively high Higgs mass measured by the LHC points to somewhat large corrections to the tree-level Higgs mass, implying larger masses for the 
squarks and gauginos.  This, in turn, increases the natural mass of the neutralinos, implying larger values closer to the TeV scale \citep{Olive}.  Additionally, given the large expected Sommerfeld enhancement from W and Z exchange for neutralinos above a few TeV, sensitive searches at larger mass scales such as that offered by VERITAS GC observations can also provide constraints on dark matter for annihilation cross sections close to the natural scale for a thermal relic ($\langle \sigma v\rangle \sim$ 3$\times$10$^{-26}$ cm$^{3}$ s$^{-1}$).

\begin{figure}[t]
\begin{center}
   \includegraphics[width=0.6\textwidth,height=70mm]{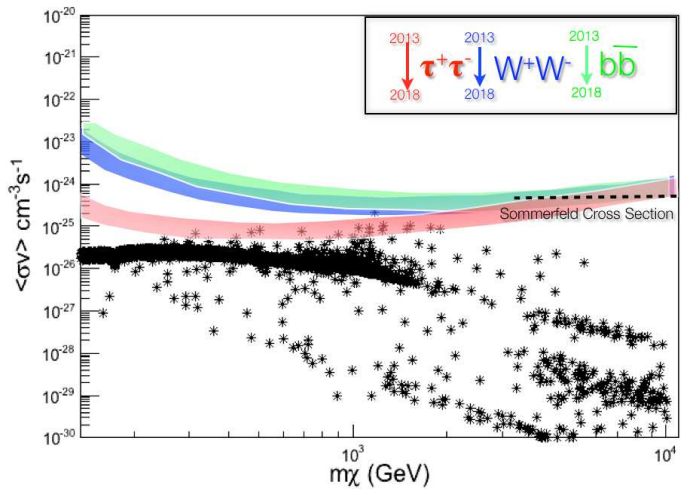}
   \caption{The predicted sensitivity for the continued VERITAS DM program. On the left is shown the predicted constraints from the dSph stacking analysis for various pure annihilation channels. The width of each colored region represents the change in predicted constraints for data accrual ending in 2013 versus 2018. }

\end{center}
\end{figure}

\subsection{Constraints From a Fully Completed VERITAS DM Program}

In the upcoming years, VERITAS will continue its dedicated indirect dark matter detection program, allocating a substantial fraction of time to DM targets which can be used in the analysis presented above. As can be seen from Figure 4, under even most the conservative estimates (i.e. no Sommerfeld or density enhancement) VERITAS will provide a considerable increase in sensitivity to both constrain and detect neutralino dark matter. Figure 4 shows the benchmark estimates for the VERITAS DM program for the predicted data accrual at the end of July 2013, and an accrual of data over an additional 5 years. One of the most important gains for the VERITAS DM programs comes from source stacking of dSph observations, allowing for a single constraint to be made from multiple observations.This technique has recently been employed by \citet{Savvas} to acquire constraints from Fermi-LAT observations of dSphs. This technique is now being applied to the VERITAS data to produce the first VERITAS dSph stacked upper limits. Over the full execution of the VERITAS DM program, this approach will yield the most exhaustive constraints on IACT dSph observations and will strongly compliment lower energy range DM constraints from Fermi-LAT. We emphasize that these predicted limits are conservative in that they do not incorporate any expected improvement in advanced analysis techniques which can offer an increase in sensitivity by a factor of 2-3.

As discussed previously, one of the chief uncertainties in deriving robust DM limits is the lack of understanding of the astrophysical density profile of DM (the J factor). As precise kinematical measurements of the dSphs soon become available \citep{Walker}, these J factors will become better constrained, yielding more precise constraints on both $\langle \sigma v\rangle $ for minimal SUSY models, as well as stronger constraints on specialized models (such as those utilizing leptophillic and Sommerfeld enhancements). 

All of the results and predictions presented above focus on the ``continuum signal" from DM annihilation. These results do not assume observation of the gamma-ray line feature,  as its contribution to the overall continuum signal is expected to be negligible. Despite this, theres has been some recent evidence derived from Fermi-LAT observations \citep{wenigerline, otherline1, otherline2} indicating a possible line signature near 130 GeV. While these results have generated considerable excitement, their grounding as evidence for DM annihilation is still in a very preliminary state and those line features  could very easily prove to be a systematic instrumental effect \citep{AAlbert}. However, the investigation of such putative line features is certainly accessible with VERITAS; the recent (Summer 2012) high-QE PMT upgrade of the instrument has extended the VERITAS triggering threshold to below 70 GeV. With comparable energy resolution to Fermi-LAT in this energy regime, VERITAS is able to contribute to either the confirmation or refutation of this line signal as it should be present in other astrophysical targets (such as the sample of stacked dSphs). Analysis towards this end is underway and will be the focus of future work.

\section{Discussion}

While the VERITAS DM program has already produced several significant results,  the most important results from the program will be forthcoming. Over the next several years, VERITAS will accrue over 1000 hours of data on a range of dark matter targets; this accrual of data will serve as the most comprehensive and far ranging of its type. Techniques such as source stacking, and refined measurement of astrophysical parameters of targets will yield both more precise and sensitive measurements. 

It is worth placing measurements of this kind in context of current and upcoming experiments across the range of direct, indirect, and collider dark matter detection techniques. Direct detection experiments such as XENON100 have made extremely sensitive measurements and have severely constrained the available parameter space \citep{XENON100}. Additionally, the LHC's dedicated SUSY searches continue to accrue new search data. Indirect detection experiments are a necessary component to these searches. Even in the scenarios where direct detection or collider searches reveal a statistically significant excess, there would be no reliable association with cosmological dark matter unless instruments making astrophysical observations are able to see an excess at the same energy over a range of astrophysical targets. Therefore, indirect IACT searches (such as the program offered by VERITAS) are not complimentary to direct and collider searches, they are $\textit{necessary}$ for the latter two to successfully characterize cosmological dark matter. In the worst case scenario that both direct and collider searches do not see a signal, IACT searches still provide an independent scan of the SUSY parameter space.

The upcoming next generation IACT, CTA \citep{CTA}, will provide the requisite sensitivity to refine and significantly improve all of the measurements discussed above. However, even in the most optimistic scenarios, CTA will not become fully operational until the early 2020s. The data accrual and analysis necessary to make precise measurements will not mature until several years after that. Therefore, increased, long term data accrual with the current generation of IACTs ( such as VERITAS) provides important coverage while Fermi-LAT, direct detection, and collider experiments are providing relevant measurements.

\bibliographystyle{apj}
\bibliography{refs}

\end{document}